\documentclass[aps,prc,twocolumn,superscriptaddress]{revtex4-1}

\usepackage{graphicx} 
\usepackage{multirow}
\usepackage{array} 

\usepackage{hyperref} 

\begin{document}

\markboth{Fission barriers heights in A$\sim$ 200 mass region}{K. Mahata}

\title{Fission barriers heights in A$\sim$ 200 mass region}

\author{K. Mahata} 
\email{kmahata@barc.gov.in}
\affiliation{Nuclear Physics Division, Bhabha Atomic Research Centre, Mumbai - 400085, INDIA}

\begin{abstract}
Statistical model analysis has been carried out for $p$ and $\alpha$ induced fission reactions using a consistent description for fission barrier and level density in A $\sim$ 200 mass region. 
A continuous damping of shell correction with excitation energy have been considered. Extracted fission barriers agree well with the recent microscopic-macroscopic model. The shell corrections at the saddle point were found to be not significant.
\end{abstract}


 
\maketitle


\section{Introduction}
Experimental determination of fission barrier height in mass A $\sim$ 200 continues to be a challenging problem. Accurate knowledge of fission barrier 
height is  vital not only to understand the heavy ion induced fusion-fission 
dynamics and  prediction of super heavy elements, but also other areas, 
such as stellar nucleosynthesis and nuclear energy applications as well.
In the actinide region, the fission barrier heights are comparable to the neutron separation energies and could be determined accurately from the measured fission excitation functions, which  exhibit a characteristic rise at the barrier energy followed by a flat plateau. In the  A $\sim$ 200 mass region, fission barrier heights are much higher than the neutron separation energies. 
Most of the measurement of fission cross sections in this mass region are performed at energies much higher than the fission barrier, where there are other open channels and a statistical description is essential.  

Although a number of studies have been made, there are still ambiguities in choosing various
input parameters for the statistical model analysis. According to the statistical model of compound nucleus decay, the probabilities of decay to different channels are governed by the transmission coefficient and relative density of states (phase space). The nuclear level density depends of the level density parameter ($a$) related to the single particle density near the Fermi surface and the available thermal energy (U). The ground state shell corrections in the nuclei around the doubly closed shell nucleus $^{208}$Pb (Z=82, N=126) are large and its damping with excitation energy has to be incorporated properly in the statistical model analysis. The nuclear level density of a shell closed nucleus shows same energy dependence at high excitation energy ($>$ 40 MeV) as that of  nuclei away from shell closure, if the excitation energy is measured from the liquid drop surface, indicating the complete washing out of the shell corrections at those energies~\cite{Ramamurthy70}. At intermediate energy the dependence is phenomenologically described in terms of energy dependent level density parameter approaching asymptotically to the liquid drop value~\cite{Ignatyuk75}. The phenomenological description of gradual damping of the ground state shell correction with excitation energy was obtained by examining the density of neutron resonances situated near the neutron threshold ($\sim$ 8 MeV). Recently, it has also been studied by measuring evaporation spectra in $^{208}$Pb region~\cite{Rout13}. 
The knowledge about the shell corrections at the saddle point in A $\sim$ 200 is obscure.


The heavy-ion induced fission excitation functions are not sensitive to the correlated variation of the fission barrier height and the ratio of the level density parameter at the saddle point to that at the equilibrium deformation ($\tilde{a_f}/\tilde{a_n}$)~\cite{Vigdor80,Ward83,Mahata03, Mahata06}. However, the pre-fission neutron multiplicity ($\nu_{pre}$) data is sensitive to this correlated variation and hence it could be used to constrain the statistical model parameters. However, the measured $\nu_{pre}$ can have dynamical contribution, which should be taken care of. 
Analysis \cite{Mahata06} of the fission and evaporation residue  cross-sections 
along with pre-fission neutron multiplicities data for $^{12}$C+$^{198}$Pt system required large shell corrections at the saddle point
and yielded fission barriers 
much smaller (13.4 MeV) than those ($\sim$ 22 MeV) obtained for 
same compound nuclei from the analysis of light ion induced reactions. In the analysis, the $\nu_{pre}$ data was corrected for a dynamical emission corresponding to fission delay of 30$\times10^{-21}$s. 

In the present article we report the statistical model analysis for $p$+$^{209}$Bi, $\alpha$+$^{184}$W, $^{206,208}$Pb, $^{209}$Bi systems using the same prescription for level density and fission barrier as in Ref.~\cite{Mahata06}. The experimental fission excitation functions are taken from Ref.~\cite{Khodai66, Zhukova77, Ignatyuk84, Moretto95, DArrigo94}.

\section{Statistical model analysis}
%
\begin{figure}[h!]
\begin{center}
\includegraphics[width=0.9\columnwidth]{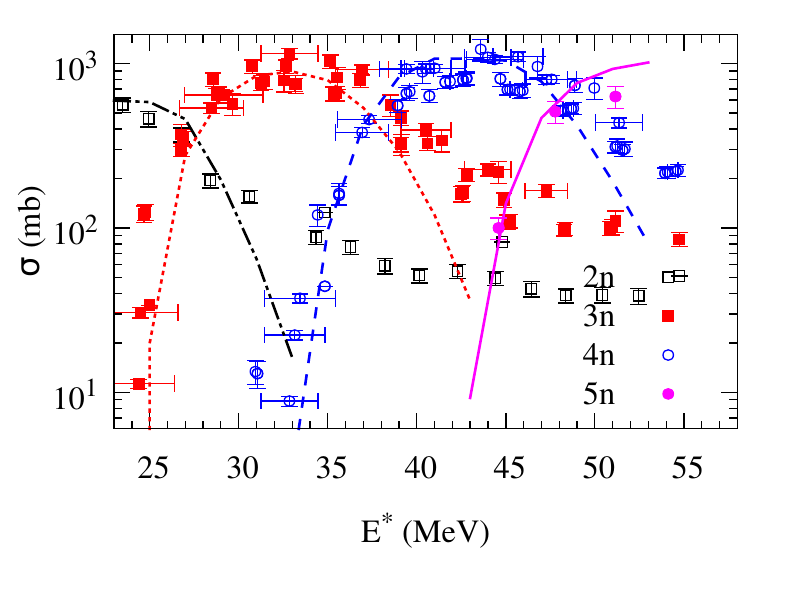}
\caption{Statistical model predictions of $xn$ cross-sections for $p$ + $^{209}$Bi are compared with the experimental data.
}
\label{fig:bimg}
\includegraphics[width=0.9\columnwidth]{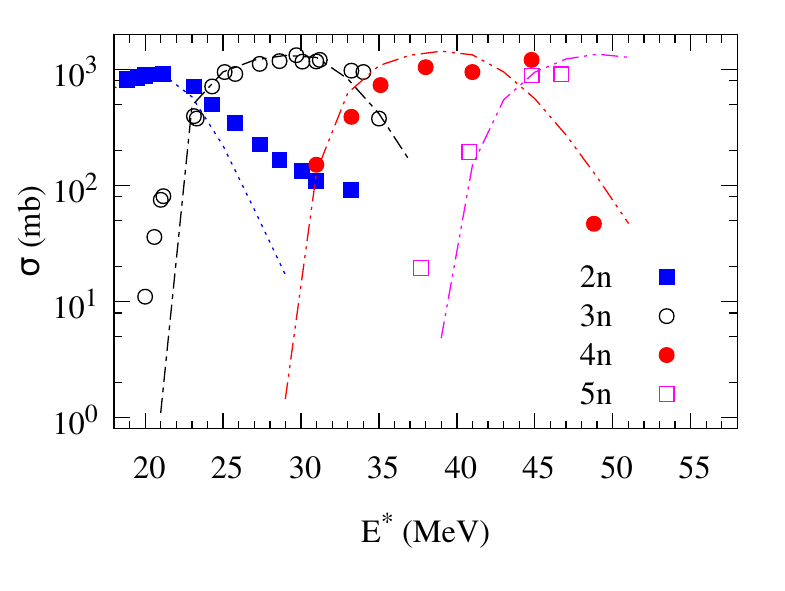}
\caption{Same as Fig.~1 for $\alpha$ + $^{209}$Bi system. 
}
\end{center}
\end{figure} 

\begin{figure}[h!]
\begin{center}
\includegraphics[width=0.7\columnwidth]{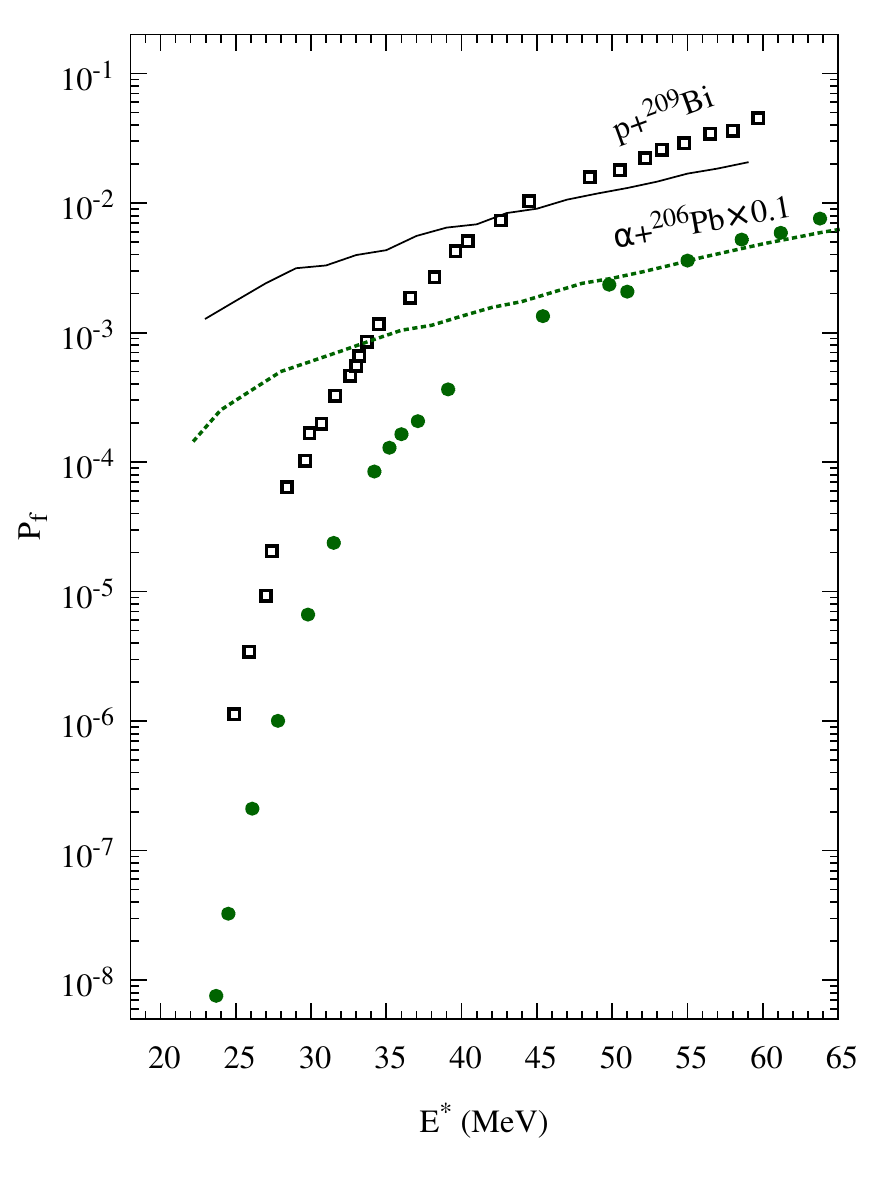}
\caption{Experimental fission probabilities  for $p$ + $^{209}$Bi and $\alpha$+$^{206}$Pb systems are compared with statistical model calculations using parameters from Ref.~\cite{Mahata06} (B$_f$(0) = 13.4 MeV) obtained from the fit to the experimental ER and fission excitation functions along with the pre-fission neutron multiplicity data for $^{12}$C+$^{198}$Pt system.}
\label{fig:widefig}
\end{center}
\end{figure} 

\begin{figure}[h!]
\begin{center}
\includegraphics[width=0.7\columnwidth]{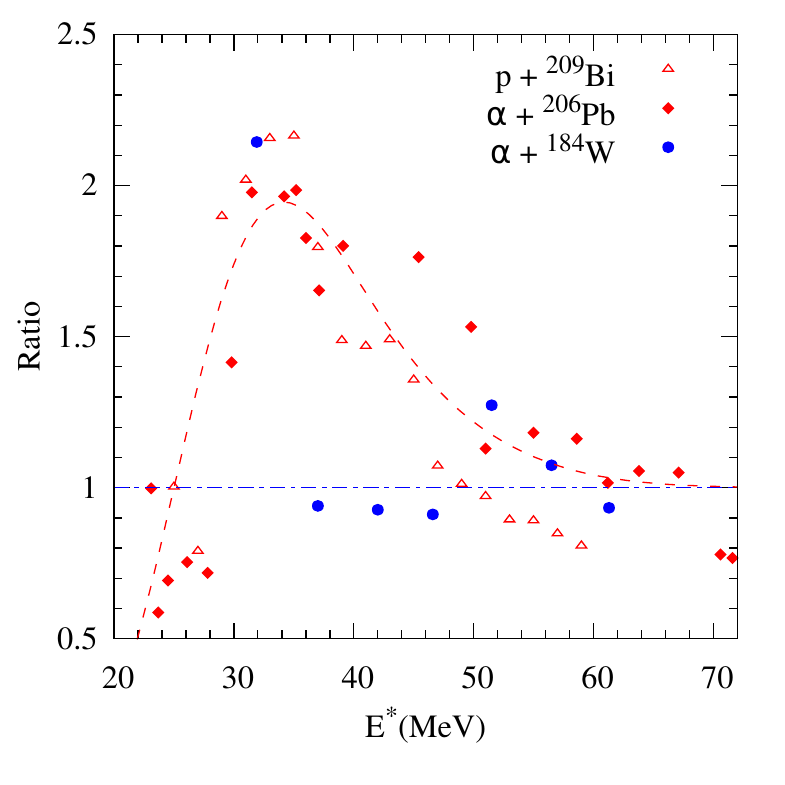}
\caption{Ratio of the experimental fission probability to the statistical model prediction using energy independent damping factor $\eta$ = 0.054. The lines are to guide the eye.}
\label{fig:widefig}
\end{center}
\end{figure} 

The statistical model analyses have been carried out using the code {\small PACE}~\cite{Gavron80} with a modified prescription for fission barrier and level density~\cite{Mahata06}. The fission barrier is  expressed as 
\begin{equation}
B_f(J) = c_f\times B_F^{RFRM}(J) - \Delta_n + \Delta_f, 
\end{equation}
where $c_f$, $\Delta_n$  and $\Delta_f$ are scaling factor to the rotating finite range model (RFRM) fission barrier~\cite{Sierk86}, shell correction at the equilibrium and shell correction at the saddle deformation, respectively. The shell corrections at the equilibrium deformations are taken from Ref.~\cite{Myers94}. 
Fermi gas level density formula has been used to calculate level densities at the equilibrium  and saddle point deformation. The excitation energy of the compound nucleus is calculated as 
\begin{equation}
U_n = E_{cm} + Q - E_{rot}(J) - \delta_p,
\end{equation}
where E$_{cm}$, Q, E$_{rot}$(J) and $\delta_p$ are the energy in the center of mass, Q-value for fusion, rotational energy and pairing energy, respectively. The Q-value for fusion as well as the particle separation energies for subsequent decays are calculated using the experimental masses~\cite{Wang12}.
The excitation energy available at the saddle point deformation is taken as $U_f = U_n - B_f(J)$. The damping of the shell correction with excitation energy is taken care by assuming energy dependent level density parameter as
\begin{equation}
a_x (U) = \tilde {a_x}[1 + (\Delta_x/U_x)(1 - e^{-\eta U_x})]
\end{equation}
with $x = n$ and $f$ corresponding to the equilibrium deformation and saddle point deformation, respectively. The asymptotic liquid drop value of the level density parameter at the equilibrium deformation is taken as $\tilde{a}_n$~= A/9~MeV$^{-1}$.  The asymptotic liquid drop value of the level density parameter at the saddle point  deformation ($\tilde{a}_f$) may be different from that of $\tilde{a}_n$ due to difference in nuclear shapes at these two cases. Hence, the ratio $\tilde{a}_f/\tilde{a}_n$ have been kept as free parameter to be decided by the fit to the data.

The spin distribution of the compound nucleus is taken as 
\begin{equation}
\sigma(\ell) = \frac{\pi(2\ell+1)}{k^2[1+\exp((\ell-\ell_{max})/\delta \ell)]}
\end{equation}
The value of the $\ell_{max}$ and $\delta\ell$ are determined from the experimental fusion cross-section ($\sigma_{fus}$) and fission fragment angular anisotropy data. Sum of the $xn$ cross-sections available in literature~\cite{Exfor} have been taken as fusion cross-section for $p$ + $^{209}$Bi system.  For $\alpha$+$^{206}$Pb system $xn$ cross-sections are not available. However, $xn$ cross-sections for $\alpha$ + $^{209}$Bi system are available~\cite{Penionzhkevich2002}. Statistical model analysis of $\alpha$ + $^{209}$Bi system taking fusion cross-section from the Bass systematics~\cite{Bass77} reproduces the experimental $xn$ cross-sections well. Hence, fusion cross-sections for  $\alpha$ induced reactions have been estimated using the Bass systematics. The value of $\delta\ell$ = 2 and 3 
reproduces the experimental fission fragment angular anisotropy data for $p$ and $\alpha$ induced reactions, respectively.

In Fig. 1 and 2, we have compared the statistical model predictions with the experimental $xn$ cross-sections for $p$ and $\alpha$ induced reaction on $^{209}$Bi target. Fusion process leads to bell shaped excitation functions for $xn$ channels and the presence of pre-equilibrium particle emission gives rise to high energy tails to these distributions. A significant contribution from pre-equilibrium particle emission to the $xn$ cross-sections will lead to over estimation of the fusion cross-section. As can be seen from the figures, the distribution of the $xn$ cross-sections could be well reproduced, except the high energy tails in the experimental distributions. We have estimated the contribution of the pre-equilibrium emission to to $xn$ cross-sections from these high energy tails. It was found that the pre-equilibrium contribution to the $xn$ cross-sections is of the order of 10\% for excitation energies below $<$40 MeV and it becomes around 20\% at 50 MeV for $p$ induced reaction. Fission following pre-equilibrium particle emission will be negligible because of the population of the target-like nuclei with much lower excitation energies than that of the compound nucleus. For the $\alpha$ induced reaction, the pre-equilibrium contribution to $xn$ cross-section is found not to be significant in the energy range considered in the present analysis.

\begin{figure}[h!]
\begin{center}
\includegraphics[width=0.7\columnwidth]{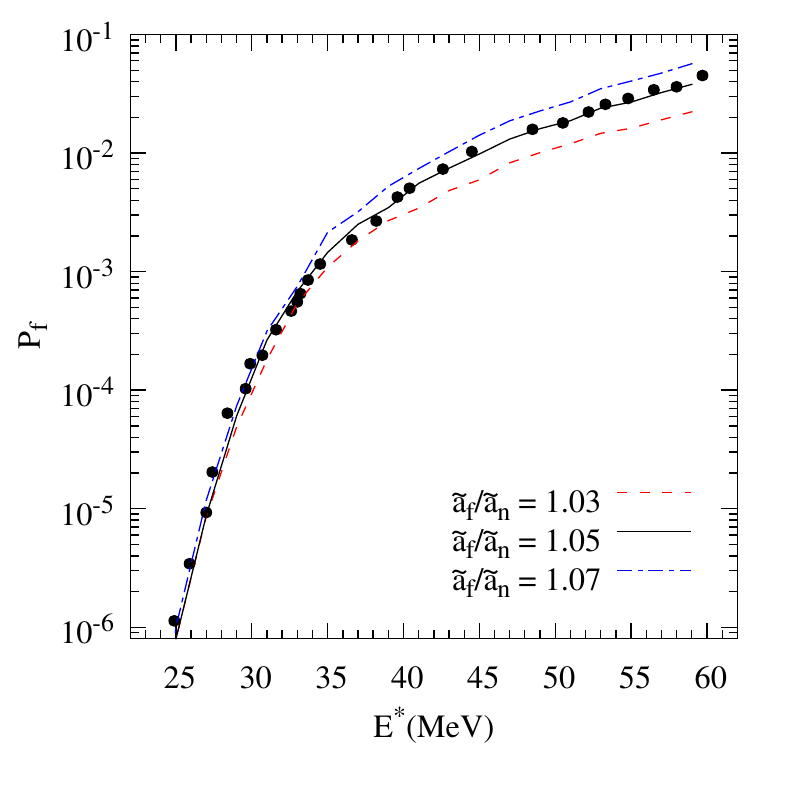}
\caption{Statistical model predictions of fission probabilities for $p$ + $^{209}$Bi system with different values of ${\tilde{a}_f}/\tilde{a}_n$ for a fixed value of B$_f$.}
\includegraphics[width=0.7\columnwidth]{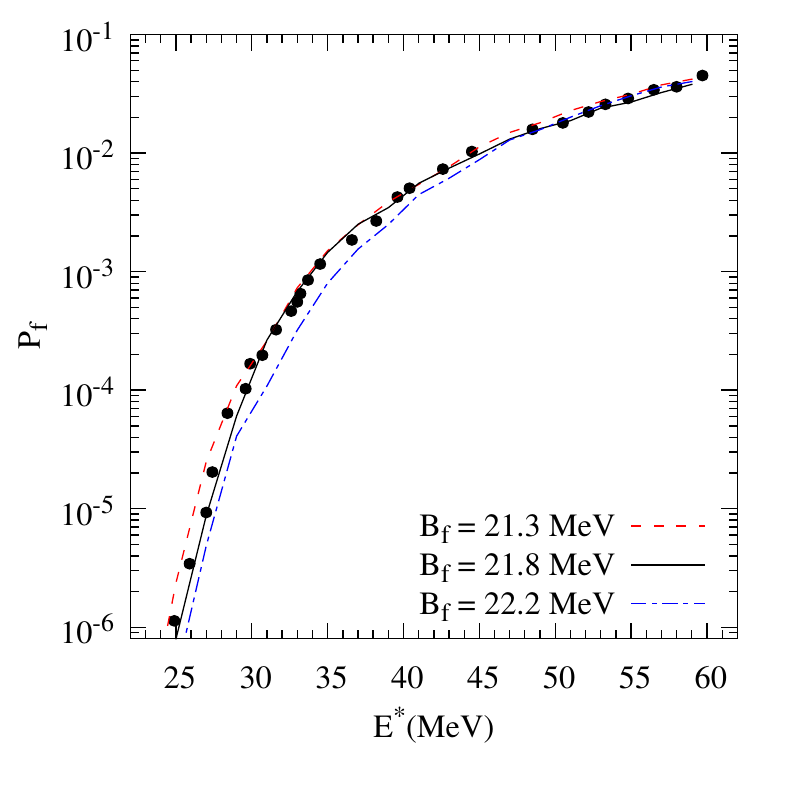}
\caption{Statistical model predictions of fission probabilities with different values of B$_f$ for $p$ + $^{209}$Bi. The values of ${\tilde{a}_f}/\tilde{a}_n$ is varied to get the best fit.}
\label{fig:widefig}
\end{center}
\end{figure}

Experimental fission probabilities ($P_f = \sigma_{fis}/\sigma_{fus}$) for $p$ + $^{209}$Bi and $\alpha$ + $^{206}$Pb systems are compared with the predictions of the statistical model using the parameters (B$_f$ = 13.4 MeV) of Ref.~\cite{Mahata06}, which fits the fission excitation functions and the $\nu_{pre}$ data for $^{12}$C+$^{198}$Pt system simultaneously. The statistical model calculation using the parameters of Ref.~\cite{Mahata06} fails to reproduce the shape of the excitation functions. Attempts to fit the excitation functions by varying the values of $\Delta_f$ and $\tilde{a}_f/\tilde{a}_n$ results $\Delta_f \sim$ 0.5 MeV. Since the resulted shell correction at the saddle point ($\Delta_f$) is within the uncertainty of the RFRM prediction and the damping of shell corrections at the saddle point is also not well known, the value of $\Delta_f$ have been assumed to be zero and the values of $c_f$ and $\tilde{a}_f/\tilde{a}_n$ have been varied to fit the excitation functions.

Initially, we attempted to fit the excitation functions using an energy independent damping factor $\eta$ = 0.054 MeV$^{-1}$~\cite{Ignatyuk75} for damping of the shell correction at the equilibrium deformation. It could not fit the excitation functions for the entire excitation energy range considered in the analysis. The ratio of the experimental fission probabilities to the predictions of the statistical model using energy independent damping factor for $p$ + $^{209}$Bi, $\alpha$ + $^{184}$W, $^{206}$Pb systems are shown in Fig. 4. While the ratio  shows a systematic deviation from unity for the $p$ + $^{209}$Bi, $\alpha$ + $^{206}$Pb ($\Delta_n$ = -10.6 MeV) systems,  no such deviation was observed for the $\alpha$ + $^{184}$W ($\Delta_n$ = -1.9 MeV) system. From this observation it was concluded that the deviation is due to the improper damping of the shell correction with excitation energy. An energy dependent shell damping factor $\eta = 0.054 + 0.002\times E^*$ is found to give better agreement with the experimental data. 

We have also studied the sensitivity of the parameters to the excitation functions. Predictions of the  statistical model with different values of $\tilde{a}_f/\tilde{a}_n$ for B$_f$ = 21.8 MeV are compared with the experimental fission probabilities for $p$ + $^{209}$Bi systems in Fig.~5. It was found that the lower part of the excitation function is not sensitive to the values of the $\tilde{a}_f/\tilde{a}_n$. Predictions of the statistical model with different values of B$_f$ for $p$ + $^{209}$Bi system are shown in Fig.~6. For each value of the fission barrier, the value of the $\tilde{a}_f/\tilde{a}_n$ has been varied to obtain the best fit. The low energy part of the fission excitation functions are found to be very sensitive to the variation of B$_f$. It should be mentioned here that pre-equilibrium particle emission will not have significant effect at the lower part of the excitation function and hence on the extracted fission barrier hight.
\begin{figure}[h]
\begin{center}
\includegraphics[width=0.7\columnwidth]{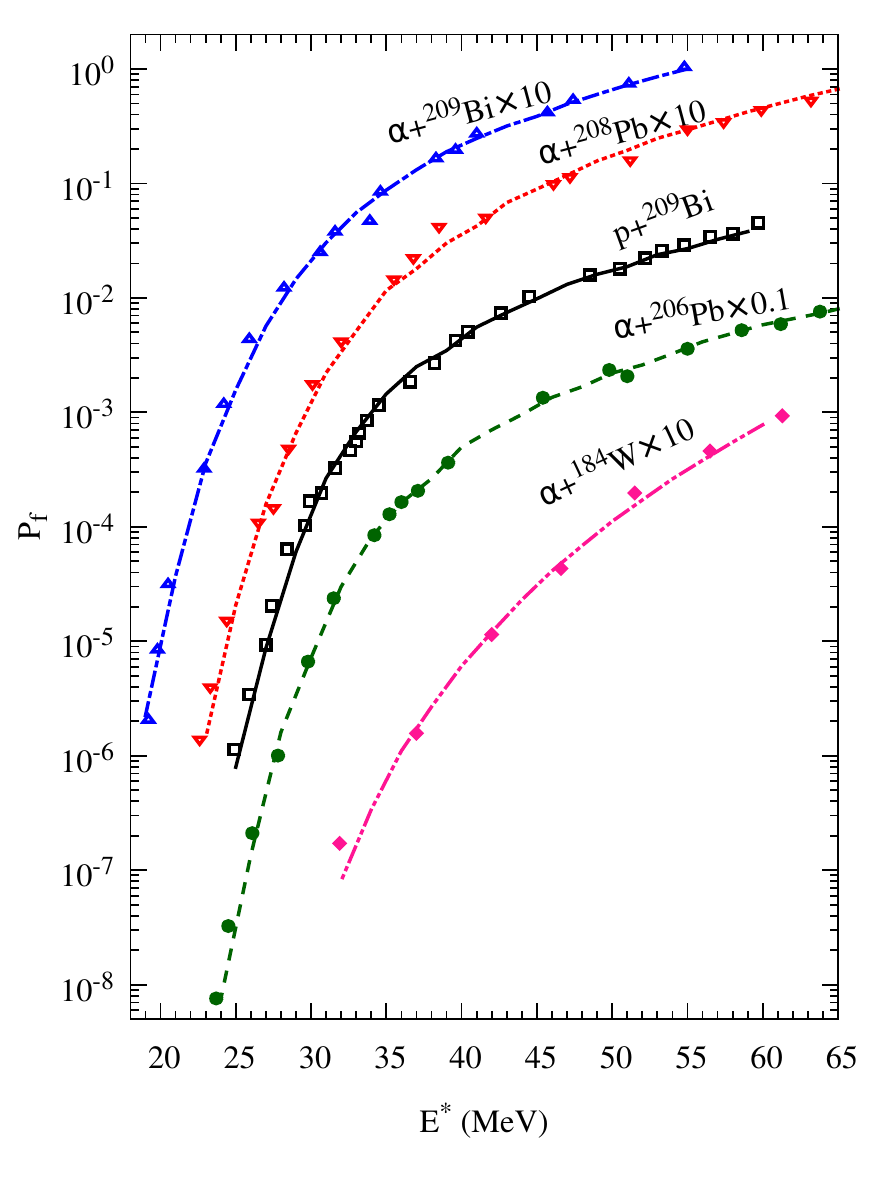}
\caption{Experimental fission probabilities for $p$ and $\alpha$ induced reactions are compared with statistical model calculations using the parameters given in the Table~1. }
\label{fig:widefig}
\end{center}
\end{figure} 

The results of the statistical model calculations are shown in Fig.~7. Best fit parameters are listed in Table~\ref{t1}.
\begin{table}[h]
\caption{\label{t1} Relevant statistical model parameters corresponding to best fits. Fission barriers extracted from the present analysis~(B$_f^{exp}(0)$) are also compared with the macroscopic-microscopic finite-range liquid-drop model~\cite{Moller09} fission barrier~(B$_f^{cal}$).}
\begin{center}
\begin{tabular}{|c|c|c|c|c|c||}
\hline 
System &$\tilde {a_f}/\tilde {a_n}$ & $c_f$ & $\Delta_n$ &  B$_f^{exp}(0)$ &B$_f^{cal}$\cite{Moller09}  \\ 
 &  &  &   (MeV) & (MeV)  \\ \hline

$\alpha$ + $^{209}$Bi $\rightarrow ^{213}$At & 1.034 & 1.04 & -7.51  & 17.8$\pm$0.2  & 18.56\\
$\alpha$ + $^{208}$Pb $\rightarrow ^{212}$Po & 1.035 & 1.06 & -8.45  & 20.3$\pm$0.3  & 20.27\\
$p$ + $^{209}$Bi $\rightarrow ^{210}$Po        & 1.042 & 1.04 & -10.62 & 21.8$\pm$0.2  & 22.14 \\
$\alpha$ + $^{206}$Pb $\rightarrow ^{210}$Po & 1.051 & 1.07 & -10.62 & 22.1$\pm$0.3  & 22.14\\
$\alpha$ + $^{184}$W$\rightarrow ^{188}$Os   & 1.082 & 1.18 & -1.89  & 24.3$\pm$0.6  & - \\
\hline
\end{tabular}
\end{center}
\end{table}

\section{Summary and conclusion}
We have carried out statistical model calculations for $p$ and $\alpha$ induced reactions to determine the fission barrier heights in A$\sim$200 mass region. Sensitivity of relevant parameters have been studied. While the low energy part of the excitation functions are found to be sensitive to the height of the fission barrier, high energy part of the excitation functions are found to be sensitive to the value of $\tilde{a}_f/\tilde{a}_n$. Effect of pre-equilibrium particle emission on the extracted fission barrier is estimated to be not significant. The shell correction at the saddle point is found to be not significant. However, the statistical model calculation without shell correction at the saddle point substantially under predict the measured pre-fission neutron multiplicity data in heavy ion fusion-fission reactions~\cite{Mahata06,Golda13}. Further investigation is required to study the statistical nature of these pre-fission neutron and contributions of  other sources (e.g. dynamical emission and near scission emission).

\section*{Acknowledgment}
The author gratefully acknowledges the suggestions and advice of Drs. S. Kailas and S. S. Kapoor.

\bibliographystyle{pramana}
\providecommand{\noopsort}[1]{}\providecommand{\singleletter}[1]{#1}%

%
%

\end{document}